\documentstyle[11pt,aaspp4]{article}

\newcommand{\bc}{\begin{center}}
\newcommand{\ec}{\end{center}}
\newcommand{\bl}{\begin{flushleft}}
\newcommand{\el}{\end{flushleft}}

\newcommand{\beq}{\begin{equation}}
\newcommand{\eeq}{\end{equation}}

\def\ms{$M_\odot$}
\def\e#1{$\times$ 10$^{#1}$}
\def\etal{et al. }
\def\ni{$^{56}$Ni}
\def\co{$^{56}$Co}

\def\ri{$R_{\rm i}$}

\def\kms{ km s$^{-1}$}

\begin{document}

\title{INSTABILITIES AND MIXING IN SN 1993J} 

\author{Kohichi IWAMOTO, Timothy R. YOUNG, Naohito NAKASATO, \\
Toshikazu SHIGEYAMA, Ken'ichi NOMOTO} 
\affil{Department of Astronomy, School of Science,
University of Tokyo,\\ Bunkyo-ku, Tokyo 113, Japan}

\author{Izumi HACHISU}
\affil{Department of Earth Science and Astronomy, College of
Arts and Sciences, University of Tokyo,\\ Meguro-ku, Tokyo 153, Japan}

\centerline{and}

\author{Hideyuki SAIO}
\affil{Department of Astronomy, Faculty of Science, Tohoku University
Sendai, Japan}

\received{20 May 1996}

\begin{abstract}

Rayleigh-Taylor (R-T) instabilities in the explosion of SN 1993J are
investigated by means of two-dimensional hydrodynamical simulations.
It is found that the extent of mixing is sensitive to the progenitor's
core mass and the envelope mass.  Because the helium core mass (3 - 4
\ms) is smaller than that of SN 1987A, R-T instabilities at the He/C+O
interfaces develop to induce a large scale mixing in the helium core,
while the instability is relatively weak at the H/He interface due to
the small envelope mass.  The predicted abundance distribution, in
particular the amount of the \ni~ mixing, is compared with those
required in the theoretical light curves and the late time optical
spectra.  This enables us to specify the progenitor of SN 1993J in
some detail.

\end{abstract}



\section{INTRODUCTION}

     SN 1993J has been identified as a Type IIb supernova (SN IIb)
from the spectral changes which show growing features of helium and
oxygen, and from the optical light curve which shows double peaks
(Wheeler \& Filippenko 1996; Baron \etal 1995 for reviews and
references therein).  These features are distinct from those of
previously known Type II supernovae (SNe II).  It was obvious that the
peculiar light curve of SN 1993J cannot be accounted for by an
explosion of an ordinary red supergiant with a massive hydrogen-rich
envelope, which produces a light curve of a SN II-P; instead, it can
be well reproduced as the explosion of a red-supergiant whose
hydrogen-rich envelope is as small as \(< \) \ 1 \ms~ (Nomoto \etal
1993; Podsiadlowski \etal 1993; Shigeyama \etal 1994; Bartunov \etal
1994; Utrobin 1994; Woosley \etal 1994; Young \etal 1995).  The
progenitor of SN 1993J is likely to have lost most of its H-rich
envelope due to the interaction with its companion star in a binary
system.

Constraints on mixing in the ejecta of SN 1993J come from both
observed spectra and photometry.  Spectroscopically, the asymmetric
profiles of the [O I] and [Mg I] emission lines at late times indicate
mixing in the ejecta (Spyromilio 1994; Wang \& Hu 1994). Shigeyama
\etal (1994) first examined two extreme cases of mixing in their light
curve modeling (i.e., with complete homogeneous mixing inside the
helium layers and without mixing) and favored mixing of \ni~ in SN
1993J.  Woosley \etal (1994) also favored mixing of \ni~ in their
light curve modeling.

     The occurrence of mixing and clumpiness in the supernova ejecta
was first confirmed in SN 1987A.  One clear confirmation is the early
detection of hard X-rays and \(\gamma\)-rays originated from the
decays of \ni~ and \co~ (e.g., Kumagai \etal 1989).  Stimulated by
these observations and theory, a number of multidimensional
simulations have been performed to show that the R-T instability
develops indeed in the ejecta of SN 1987A (Arnett \etal 1989; Hachisu
\etal 1990, 1992; M\"uller \etal 1991; Den \etal 1990).  It was found
that the instability is weak at the He/C+O and Ni/O interfaces but
strong at the H/He interface, which can be expected from the large
helium core mass \( \sim \) 6 \ms~ and the massive hydrogen-rich
envelope \( \sim 10 \) \ms.  In SN 1987A, the large scale mixing of
nickel was induced by falling spikes of hydrogen which penetrated into
deep inner layers.

     Later, it has been recognized that the occurrence of mixing and
clumping in the supernova ejecta can occur also in other types of
supernovae and that the effect on their light curves can provide
interesting diagnostics of the internal structure of the progenitors.
The extent of mixing and clumpiness affect the light curve and
spectra.  A number of hydrodynamical simulations have shown that the
R-T instabilities arise in the Type II-P explosions of red-supergiants
(Herant \& Woosley 1994; Shigeyama \etal 1996) and helium star models
of Type Ib supernovae (Hachisu \etal 1991, 1994) as in SN 1987A.

     These studies indicate the importance of multidimensional
simulations of instabilities in supernova ejecta.  This is
particularly interesting for SN 1993J because its presupernova
structure is suggested to be quite different from other supernovae.
Conversely, we may strongly constrain the progenitor's mass and
structure because the R-T instabilities are sensitive to the density
structure of the progenitor.  Despite such importance, no such
simulation has been conducted before.  Therefore, we have carried out
two dimensional hydrodynamical simulations for the R-T instability in
the ejecta of SN 1993J to examine the extent of mixing and the
resultant abundance distributions quantitatively.  In parallel, we
have calculated bolometric light curves for several explosion models
with various extent of mixing and compared them with the observations.
These results would enable us to constrain the progenitor model, which
would be useful to understand the still debated presupernova
evolution.

     In the next section, the presupernova models for SN 1993J are
described. Results of the linear stability analysis and the 2D
hydrodynamical simulations are shown in \S3 and \S4, respectively.
The effects of \ni~ mixing on the optical light curve are discussed in
\S5.  Finally we summarize some constraints on the explosion models.

\section{HYDRODYNAMICAL MODELS}

\subsection{Presupernova Models of the Progenitor}

     The progenitor models are constructed following Shigeyama \etal
(1994).  We use the presupernova helium cores of masses \(M_\alpha =
3.3 M_\odot \) and \( 4 M_\odot \) (Nomoto \& Hashimoto 1988), which
are evolved from the main-sequence stars of \(M_{ms} = 13 M_{\odot}\)
and \( 15 M_{\odot}\), respectively.  The adopted main-sequence masses
are consistent with the inferred progenitor's luminosity (Hashimoto,
Iwamoto, \& Nomoto 1993).  We assume that the star retained only a
small mass H-rich envelope after a large amount of mass loss due to
the merging with a small mass companion star (Nomoto \etal 1995,
1996).  The hydrostatic and thermal equilibrium H/He envelope models
are constructed for various masses \(M_{\rm env}\), radii \ri, and the
helium abundance $Y$ to smoothly fit to the helium core.  The model
parameters are summarized in Table 1.  The density distributions for
3H11, 3H41, 4H13, and 4H47 are shown in Figure 1.

     These models have large helium abundances in the envelope.  Such
an enhancement of helium is expected from the gradient of helium
abundance in the deepest layers of the envelope (see, e.g., Saio,
Nomoto, \& Kato 1988).  If most of the hydrogen-rich envelope is
removed, the helium-rich layer is exposed as seen in Yamaoka \etal
(1991) and Nomoto \etal (1991); they have calculated mass loss from
the initially 13 \ms~ star, which reduces the envelope mass down to
0.3 \ms~ and enhances the average helium mass fraction up to \(\sim\)
0.5 at the termination of Roche lobe overflow (see also Woosley \etal
1994).

\subsection{Hydrodynamics of Explosion} 

     We simulate the core collapse-induced explosion by depositing
thermal energy at the mass cut (a partition between neutron star and
ejecta) generating a strong outward shock wave.  The deposited energy
is set to produce the kinetic energy of explosion $E$ = 1 \e{51} erg.
The mass cut is chosen to produce 0.1 \ms\ of \ni. The propagation of
the shock wave and hydrodynamics of the ejecta are calculated with a
one dimensional Lagrangian PPM code (Shigeyama \etal 1991).
Nucleosynthesis yields behind the shock wave are nearly the same as
obtained in the 3.3 \ms~ and 4.0 \ms~ He core models (Shigeyama
\etal 1990; Iwamoto \etal 1994; Thielemann, Nomoto, \& Hashimoto
1996).

     Figure 2 shows the evolution of the density and pressure profiles
in the ejecta of 4H13 and 3H11, respectively.  When the shock wave
passes through the He/C+O interface, the ejecta around the shock front
undergoes a strong deceleration, which forms an inwardly propagating
reverse shock.  This stage corresponds to the profiles at \( t = 10 \)
sec in Figure 2.  Between the two shock waves, there appears a region
with pressure inversion.  Then the R-T instability is expected to
arise around the interface due to the opposite signs of the density
and pressure gradients.  When the shock wave arrives at the H/He
interface, it is strongly decelerated again in the same manner as it
passed the He/C+O interface.  This is shown in the profiles at \( t =
1000\) sec in Figure 2.

     This complicated behavior of hydrodynamics, the alternate
acceleration and deceleration, can be explained by behavior of the
self-similar solutions of the blast shock in spherically expanding gas
with a power law density distribution \( \rho \propto r^{-n} \) (Sedov
1959; Chevalier 1976).  In these solutions, the forward shock wave is
accelerated for \(n > 3 \), while the shock is decelerated for \(n < 3
\) (M\"uller et al. 1989; Herant \& Woosley 1994). Figure 3 shows \(
\rho \ r^{3} \) against the enclosed mass \(M_{r}\) for the initial
density profiles of 4H13 and 3H11, respectively.  It is seen that
there are two distinct regions where the shock wave should be
decelerated and the R-T instability would grow.

\section{LINEAR STABILITY ANALYSIS}

     In this section, we present an estimate of the R-T growth by the
linear stability analysis.  Our method is basically the same as that
used for SN 1987A by Ebisuzaki et al.(1990), Benz \& Thielemann
(1991), and M\"uller et al. (1991).  In the classical case that two
homogeneous fluids are stratified in the gravitational field, a linear
growth rate is given as (Chandrasekhar 1981)

\begin{equation}
G_{RT}^{2} = \frac{\rho_{+}-\rho_{-}}{\rho_{+}+\rho_{-}} kg ,
\end{equation}

\noindent
where \(\rho_{+}\) and \(\rho_{-}\) are the densities in the upper and
lower layers, respectively, and \(k\) is the wavenumber of the
perturbation, \(g\) the acceleration due to the gravity. This formula
can be applied to the shocked ejecta of a supernova by replacing \(g\)
with the effective gravity \(g_{\rm eff}\) defined as

\begin{equation}
g_{\rm eff} = - g - \frac{1}{\rho} \frac{d P}{d r}
\simeq - \frac{1}{\rho} \frac{d P}{d r}.
\end{equation}

The pressure gradient term dominates in the strongly decelerated blast
wave and the stellar gravity is negligibly small in the expanding
ejecta.  Equations (1) and (2) show that the ejecta gets R-T unstable
if the density and pressure gradients have opposite signs,

\begin{equation}
\frac{d P}{d r} \frac{d \rho}{d r} < 0 . 
\end{equation}

We calculate the growth factor of the amplitude at each layer,
\(\zeta/\zeta_{0}\), by integrating the growth rate at each stage of
spherically symmetric explosion models as

\begin{equation}
\frac{\zeta}{\zeta_{0}} = \exp \left[ \int_{0}^{t} 
Re( G_{RT} (t^{'})) dt^{'}
\right], 
\end{equation}

\noindent
where \( \zeta_{0}\) is the initial amplitude of the perturbation and
\(Re(G_{RT})\) indicates the real part of the growth rate.  As a
linear growth rate, we take that of the fastest growing mode with the
wavelength of about pressure scale height.  The formula of the growth
rate for the continuous density distribution (Chandrasekhar 1981)
reduces to the following expression based on the above assumption:

\begin{equation}
G_{RT}^{2} \simeq - \frac{1}{\gamma} \frac{c_{s}^{2}}{1+\pi^{2}}
\frac{1}{H_{\rho} H_{P}}.
\end{equation}

\noindent
Here \(\gamma\) is the adiabatic index, \( c_{s} \) the adiabatic
sound velocity, and \(H_{\rho}\) and \(H_{P}\) are the density and
pressure scale heights, respectively, which are defined as

\begin{equation}
H_{\rho} = \frac{d \ r}{d \ {\rm ln} \rho}, \hspace{1cm} 
H_{P} = \frac{d \ r}{d \ {\rm ln} P}.
\end{equation}

     Note that the linear stability analysis is not valid after the
instability develops into the nonlinear phase.  However, it is still
meaningful to see qualitatively which regions of the ejecta are
unstable.  We calculate the R-T growth by equation (4) in parallel
with one dimensional hydrodynamical simulations.  Figure 4a shows the
amplitude \(\zeta/\zeta_{0}\) versus enclosed mass for both 3H41 and
4H47 at $t$ = 1 day after the explosion.

There appear three distinct unstable regions: (1) the He/C+O interface
around $M_r$ = 1.8 \ms~(3H41) and 2.1 \ms~(4H47), (2) the H/He
interface around $M_r$ = 3.3 \ms~(3H41) and 4.0 \ms~(4H47), and (3)
the surface region near the density inversion as seen in Figure
1. These regions are coincident with those expected from the behavior
of \( \rho r^{3}\) in Figure 3. For 3H11 and 4H13, models with a
smaller mass envelope, the growth is almost the same as for 3H41 and
4H13, respectively, but slightly different due to the different
envelope structure (Fig.4b).

     Figures 4a and 4b also show that at the He/C+O interface the
growth is somewhat larger for 3H models than for 4H models.  This is
because the mass ratio between the He layer and the C-Ni core is
larger in the 3.3 \ms~ He core than in the 4 \ms~ core so that the
C-Ni core undergoes larger deceleration (Hachisu \etal 1991).

\section{2-D SIMULATIONS}

     We carry out two dimensional hydrodynamical simulations to follow
a nonlinear growth of the R-T instabilities and mixing.  The linear
stability analysis indicates that the instability would grow in some
particular regions, i.e, at the He/C+O interface, the H/He interface,
and near the surface.  The radius at the He/C+O interface is about
\(\sim 1 \ {R}_\odot\) for both progenitor models, being two orders of
magnitudes smaller than that of the H/He interface.  When the shock
wave reaches the H/He interface, the instability at the He/C+O
interface has already ceased to grow.  For these reasons, we examine
the instabilities in the He core and near the core/envelope boundary
in separate calculations to avoid a significant numerical diffusion
due to unnecessary rezoning.

     Our code is a standard third order TVD
(Total-Variation-Diminishing) scheme with a preprocessing flux-limiter
using the Roe's approximate Riemann solver (Hachisu \etal 1992, 1994).
It has second order accuracy in time by using two-step time
integration. We make several improvements in treating the advection of
the chemical composition of the fluid.  The advection equations of the
mass fractions are solved as parts of the extended Roe matrix, which
enables us to treat them consistently with the Euler equations
constraining the sum of the mass fractions to be unity.  This new
method is advantageous for accurately obtaining the extent of mixing.

     We use an equation of state with a constant adiabatic index 4/3,
which is a good approximation here since the ejecta remains in the
radiation dominant phase because of high temperatures.  We use
cylindrical coordinates with 513 grid points in each axis, $R$ and
$Z$.  Several rezonings are still necessary even if the mixing in the
inner and outer part is studied separately.  When the shock wave
reaches just inside the numerical boundary, we double the mesh size
and remap all the flow variables onto the new grid system.  Each
physical quantity \( q(2i-1,2j-1) \) in the old grid is projected to
\( q(i,j) \) in a quarter of the new computational area.  The initial
profiles are inserted in the remaining part.

     At the beginning of calculation, we apply random perturbations to
the velocity field interior of the shock front (Hachisu \etal 1994).
The latitudinal angle is divided into \(n\) pieces and the velocity
field is perturbed in each area of \(\pi/n\) angles as
  
\begin{equation}
v_{r} = v_{0} ( 1+ \epsilon(2 {\rm rmd}([n\theta/\pi])-1)),
\hspace{1cm} v_{\theta} =0,
\end{equation}

\noindent
where rmd(\(i\); integer) is a sequence of uniform random numbers in
the range from 0 to 1 and [\(x\)] denotes the maximum integer that
cannot exceed \(x\).  In the present study, the amplitude of the
perturbation is assumed to be \( \epsilon = 0.05 \) in all cases.  The
extent of mixing is expected to be independent of mesh resolutions
with this large amplitude of initial perturbation (Hachisu \etal 1992,
1994).

     Figures 5a and 5b show the density contour near the He/C+O
interface for 3H41 (at \(t = 174 \) sec) and 4H47 (at \( 5.34 \times
10^{4} \) sec), respectively.  Each contour is drawn with a
logarithmically equal interval.  It is seen that mushroom-like fingers
are developed in the He core.  This clearly shows that the R-T
instability has grown to its nonlinear stage and a large scale mixing
of the elements occurred in the velocity space within the He core.

     Figures 6a and 6b show the density contour in the outer part near
the H/He interface for 3H41 (at \(t = 4 \times 10^{4} \) sec) and 4H47
(at \( 5 \times 10^{5} \) sec), respectively.  The R-T instability
also grows due to the steep density gradient above the
interface. However, the finger-like structures near the bottom of the
H-rich envelope are not developed so much as in the inner core region
so that the extent of mixing is relatively small. This is due to the
small hydrogen-rich envelope mass and thus a weak deceleration.

Near the He/C+O interface in the core, the R-T growth tends to be
larger for smaller He core mass as discussed for the linear stability
analysis.  Such a dependence of mixing on the structure of the
progenitors has been shown by Hachisu \etal (1990, 1991, 1994) for
their models of SN 1987A and Type Ib supernovae.  For SN 1993J, we
expect that the R-T instability at the H/He interface is much weaker
than in SN 1987A because of the smaller envelope mass.  On the other
hand, the density gradient in the He core of SN1993J is steeper than
that of SN 1987A because of the smaller core mass.  Therefore, we
predict that the He/C+O interface in SN 1993J is the most unstable
region and a large scale mixing would be induced in the He core.

     We calculate averaged radial distributions of the chemical
compositions in the ejecta from the results of the two dimensional
simulations.  Figures 7 and 8 show the resultant abundance
distributions against $M_{r}$ for 3H41 and 4H47, respectively.  It is
seen that \(^{56}\)Ni is mixed out to $M_{r} =$ 1.0 \ms\ for 3H41 but
only to $M_{r}$ = 0.5 \ms\ for 4H47.  Mixing of carbon and oxygen into
the He layer is significant for both cases.  Hydrogen is not largely
mixed down to the He core for both cases due to the small envelope
mass.  If the mass of the hydrogen-rich envelope is smaller (as in
3H11 and 4H13 in Table 1.), the expanding core undergoes smaller
deceleration so that the velocities of H, He, O, and Fe would be
higher.  Comparisons of the observed expansion velocities of H, He, O,
and Fe with those predicted by the models will be made in \S6.

\section{OPTICAL LIGHT CURVE AND MIXING}

     As described in \S1, the optical light curve of SN 1993J showed a
unique behavior, with neither a clear plateau nor a monotonic decline.
It rapidly declined after the first maximum, then rose to the second
peak in 10-15 days, and finally followed by a gradually declining
tail.  The basic feature of this peculiar light curve can be accounted
for with the Type IIb model, namely, the explosion of a red supergiant
whose hydrogen-rich envelope is quite thin.

     The light curve analysis of SN1993J has shown that the progenitor
radius and mass are the main parameters that determine the shape of
the light curve (e.g., Shigeyama \etal 1994; Woosley \etal 1994; Young
\etal 1995).  These studies have also noted some effects of \ni~
mixing on the light curve shape.  In the earlier sections, we have
shown that \ni~ is indeed mixed into the He layer but the degree
of mixing depends on the progenitor's mass.  Here we present more
detailed study of the optical light curve in order to determine the
amount of mixing of \ni~ and examine whether the plausible model for
SN 1993J can be identified.  More details as well as the dependence on
other parameters will be presented in Young \etal (1996).

     We calculate bolometric light curves for 3H11, 3H41, and 4H47
with several changes in the \(^{56}\)Ni distribution.  The light curve
code is the same as used in Young \etal (1995), which is the
flux-limited radiative transfer code assuming LTE.  Figures 9 - 11
compare the theoretical light curves of these models with the observed
bolometric light curve (Richmond \etal 1994; Lewis \etal 1994) with a
distance to M81 of 3.63 Mpc (Freedman \etal 1995).  In each of the
figures, three degrees of mixing are used; for example, "Mix 0.2 \ms"
implies that \ni~ is uniformly mixed from the center to $M_r$ = 0.2
\ms.  The explosion energy and \ni~ masses are held at $E$ = 1 \e{51}
ergs and 0.1 \ms, respectively, for all the models.

     Generally, the first peak in the light curve is produced by the
radiation field established in the shock heated H/He envelope, while
the second peak is formed by diffusive leak of the deposited energy
due to the radioactive decay of \ni~ and \co.  Finally the light curve
enters an exponential declining tail due to the \(^{56}\)Co decay.

     The most obvious feature that the mixing effects is the tail of
the light curve.  The declining rate of the tail is faster for more
extensive mixing and for smaller mass He core, since the ejecta is
more transparent to \(\gamma\)-rays.  The no mixing case for 3H11 has
a brighter tail and a shallower slope and is brighter than the
observations.  On the contrary, mixing to 1.0 - 1.5 \ms~ for 3H41
gives a tail that is too steep to be compatible with observations.

     The effects of mixing on the light curve shape around the dip at
day 20 and the second peak are also seen.  With more extensive mixing,
the heating effect of radioactive decays starts to appear earlier so
that the dip is shallower and the second peak is reached earlier.
These effects are relatively small for 3H11 but much more important
for 3H41 and 4H47 due to the thicker envelope.  For 3H41 a critical
mixing mass is found to be out to $M_r$ = 1 \ms, with which the fit at
20 days is improved.  However mixing to this extent gives a tail that
is too steep as seen in Figure 10.  For 4H47, the tail is reproduced
well with the mixing up to 1.8 - 2.2 \ms, and the mixing out to 2.2
\ms~ gives a good fit to the dip at 20 days.

     To summarize, the observed light curve shape around the dip, the
second peak, and its tail are well reproduced with the model 4H47 if
mixing extends to $M_r \sim$ 2 \ms.  For 3H11 and 3H41, the mixing
produces better fit to the tail (solid line) but tends to form a dip
at day 20.  Improvement of the radiative transfer code might improve
the fit at the dip for 3H models.

\section{DISCUSSION}

      We compare the calculated abundance distributions as a function
of $M_{r}$ (measured from the bottom of the ejecta) (Figures 7 and 8)
and the expansion velocity (Figures 12 -- 15 ) with the degree of
mixing required from the light curve modeling and the spectroscopic
observations of SN 1993J.

     After the expansion becomes homologous ( \(v \propto r\) ), the
expansion velocity of each layer remains constant so that it can be
directly compared with the observed velocities of various elements.
In the hydrodynamical models, the expansion velocity \(v_{\rm exp}\)
depends on the explosion energy \(E_{\rm exp}\) and the total mass of
the ejecta \(M_{\rm ej}\) as

\begin{equation}
 v_{\rm exp} \propto E_{\rm exp}^{1/2} M_{\rm ej}^{-1/2}. 
\end{equation}

We have tried another 2-D simulation for 3H41 with a different
explosion energy of \( E_{\rm exp} = 0.6 \times 10^{51}\) erg and
found little difference in the degree of mixing in the $M_{r}$ space
compared with the case of \( E_{\rm exp} = 1.0 \times 10^{51}\) erg.
Therefore, we can safely assume that the velocity of each element in
our models scales to the explosion energy as in equation (8).

     Houck and Fransson (1996) have analyzed the spectra of SN 1993J
and find that the observed spectra are well reproduced by their
synthetic spectra if (1) some iron extends to at least 3000 \kms~ in
the velocity space, (2) $\sim$ 0.5 \ms~ oxygen occupies 1000 - 4000
\kms, and (3) the bulk of hydrogen lies between 8500 - 10,000 \kms.
For the hydrogen velocity, however, the minimum velocity has been
determined to be $v_{\rm H min} \sim$ 7500 \kms~ from the inner edge
of the H\(\alpha\) line profile at late times (Patat, Chugai \&
Mazzali 1995).

The coexistence of oxygen and iron at 1000 - 3000 \kms~ implies the
mixing of O and Fe in the velocity space, if O and Fe are separated at
first. Our calculations show that even the larger mass 3H model, 3H41,
has the maximum velocity of oxygen as high as 6500 \kms.  Thus, as for
the oxygen velocity, 3H models are not preferable. The maximum
velocity of oxygen in 4H47 is \(\sim\) 4300 \kms~, which is relatively
agreeable.

     The maximum velocity of \ni~ (decaying eventually to Fe) is
sensitive to the He core mass, i.e., $v_{\rm Ni} \sim$ 6000 km
s\(^{-1}\) for 3H models and 3000 km s\(^{-1}\) for 4H models.  This
is because the He core with a smaller mass expands faster and
undergoes more extensive mixing as discussed in \S3.  Compared with
the observations, $v_{\rm Ni}$ in 3H models is too fast.  In 4H
models, $v_{\rm Ni}$ \(\sim\) 3000 \kms~ is marginally acceptable but
a little too low.

     The minimum velocity of hydrogen, $v_{\rm H min}$, depends mainly
on the mass of the H/He envelope, since the effect of mixing at the
core/envelope boundary is not significant.  For 3H11 (3H41) and 4H13
(4H47), $v_{\rm H min} \sim$ 8500 (7000) \kms~ and 9000 (6300) \kms,
respectively.  Thus, if we adopt $v_{\rm H min} \sim$ 7500 \kms~ from
the late time H$\alpha$ profile, the envelope mass of $\sim$ 0.3 - 0.4
\ms~ would be more consistent with the observed hydrogen velocities
than $\sim$ 0.1 \ms.

     We find in \S4 that the calculated light curves give the best fit
to the observations if \ni~ is mixed up to $M_r \sim$ 2.2 \ms~ for
4H47 and to 0.5 - 1.5 \ms~ for 3H41 and 3H11.  However, in our
simulations of the R-T instabilities, \ni~ is mixed up to $M_r \sim$
0.5 \ms~ for 4H models and to 1.0 \ms~ for 3H models.  Thus the
calculated degree of \ni~ mixing is too small in 4H models, while
being consistent with the light curve modeling in 3H models.  We
should note, however, that the agreement between the calculated and
observed light curves is better for 4H47 than for 3H41.

     The determination of oxygen mass from the late time spectra would
be another way to determine the progenitor's mass.  Houck \& Fransson
(1996) estimated that the oxygen mass is $\sim$ 0.5 \ms.  In the
explosion models, the produced oxygen masses are 0.21, 0.42, and 1.5
\ms~ for the He cores of 3.3, 4, and 6 \ms, respectively. (Thielemann
\etal 1996). Thus the 4 \ms~ core model gives the more consistent
oxygen mass.

     The inconsistencies of the mixing in the 3H and 4H models with
the spectroscopic and photometric observations can be reconciled as
follows.

(1) For 3H models, the fast decline of the tail may be improved if the
explosion energy is as low as 0.8 \e{51} ergs.  This is also suggested
>from the X-ray light curve analysis (Suzuki \& Nomoto 1995).

(2) For 4H models, to be consistent with the observed light curve, the
mixing of \ni~ should be much more extensive than that occurs in the
2D simulation.  Such a large scale mixing would require an extremely
large initial perturbations, which might be due to the
neutrino-induced R-T instabilities (Herant, Benz, \& Colgate 1992;
Burrows, Hayes, \& Fryxell 1995; Janka \& M\"uller 1996), the
convective oxygen shell burning just before the collapse (Bazan \&
Arnett 1995), or the spiral-in of a companion star in the common
envelope scenario (Nomoto \etal 1995, 1996).

The instabilities in the H-rich envelope might be more extensive than
our models show.  An additional perturbation, such as an asymmetric
structure (H\"oflich 1995) due to the spiral-in of the companion star
into the envelope (Nomoto \etal 1995, 1996) would cause more mixing.
It is important to know whether the density distribution in the H-rich
envelope is largely changed due to the possible development of the
above instabilities, because the X-ray light curves (Zimmermann \etal
1994; Kohmura \etal 1994) has been found to be sensitive to the
envelope structure of the ejecta (Suzuki \& Nomoto 1995; Fransson,
Lundqvist, \& Chevalier 1996).

\section{CONCLUSIONS}

We have investigated the Rayleigh-Taylor instabilities in the ejecta
of SN 1993J with a linear analysis of spherically symmetric explosion
models and with a two-dimensional hydrodynamical simulations. We find
the following conclusions.

1. The instability at the He/C+O interface develops to induce a large
scale mixing because of the relatively small He core mass.

2. The instability at the H/He interface is weak because of the small
hydrogen-rich envelope mass.

These features (1,2) are in contrast to SN 1987A which had the more
massive He core and the envelope.  The extent of mixing of heavy
elements (Ni and C+O) is sensitive to the core mass.  For the smaller
core mass, the R-T instability is stronger and causes more extensive
mixing due to the smaller mass ratio between the core and the He
layer.

3. The optical light curves are calculated with a parameterized degree
of mixing.  The oberved light curve is well reproduced if substantial
amount of \ni~ mixing occurs.

4.  The calculated abundance distributions of the ejecta against the
expansion velocity are compared with the observed velocities of Ni, O,
and H.  The model with the 3.3 \ms~ He core and the hydrogen-rich
envelope of 0.3-0.4 \ms~ can well reproduce the observational feature
of SN 1993J, if the explosion energy is as low as $\sim$ 0.8 \e{51}
ergs.  The model with the 4 \ms~ He core and the $\sim$ 0.5 \ms~
envelope is also a good alternative, if \ni~ is more extensively mixed
than our present calculations, possibly due to much larger initial
perturbations.

\bigskip

This work has been supported in part by the grant-in-Aid for
Scientific Research (05242102, 06233101) and COE research (07CE2002)
of the Ministry of Education, Science, and Culture in Japan, and the
fellowship of the Japan Society for the Promotion of Science for
Japanese Junior Scientists (4227).  The computation was carried out on
Fujitsu VPP-500 at the Institute of Physical and Chemical Research
(RIKEN) and the Institute of Space an Astronautical Science (ISAS),
and VPP-300 at the National Astronomical Observatory in Japan (NAO,
Tokyo).

\newpage

\centerline{Figure Captions}

\bigskip\noindent
Fig. 1 -- Density structures of the progenitor models at the onset of
collapse.

\bigskip\noindent 
Fig. 2 -- Changes in the density and pressure profiles for (a) 4H13
and (b) 3H11. Each label indicates the time after the explosion; {\bf
1}: 0 sec, {\bf 2}: 10 sec, {\bf 3}: 100 sec, {\bf 4}: 1000 sec, {\bf
5}: 10,000 sec.

\bigskip\noindent 
Fig. 3 -- Plots of \( \rho r^{3} \) against enclosed
mass \(M_{r}\)/\(M_{\odot}\) for 4H13 (upper) and 3H11 (lower).

\bigskip\noindent 
Fig. 4 -- The Rayleigh-Taylor growth as a function of enclosed mass
\(M_{r}\)/\(M_{\odot}\) (a) \ for 3H41 (upper) and 4H47 (lower), and (b)
for 3H11 (upper) and 4H13 (lower).

\bigskip\noindent
Fig. 5 -- Density contours for (a) 3H41 (174 sec) and
(b) 4H47 (149 sec).

\bigskip\noindent
Fig. 6 -- Density contours for (a) 3H41 (5.34\(\times 10^{4}\)
sec) and (b) 4H47 (3.95\(\times 10^{4}\) sec).

\bigskip\noindent 
Fig. 7 -- The averaged radial distribution of several elements as a
function of \(M_{r}\) measured from the bottom of the ejecta for 3H41. 
Shown are the mass fractions of \(^{56}\)Ni (solid), C+O (dotted),
He (dashed), and H (long-dashed).

\bigskip\noindent 
Fig. 8 -- Same as Figure 7 but for 4H47.

\bigskip\noindent 
Fig. 9 -- Calculated bolometric light curves for 3H11 as compared with the 
observations of SN 1993J.

\bigskip\noindent 
Fig. 10 -- Same as Figure 9 but for 3H41.

\bigskip\noindent 
Fig. 11 -- Same as Figure 9 but for 4H47.

\bigskip\noindent
Fig. 12 -- The averaged radial distribution of several elements as a
function of the expansion velocity for 3H11. Shown are the mass
fractions of \(^{56}\)Ni (solid), C+O (dotted), He (dashed), and
H (long-dashed).

\bigskip\noindent
Fig. 13 -- Same as Figure 12 but for 3H41.

\bigskip\noindent 
Fig. 14 -- Same as Figure 12 but for 4H13.

\bigskip\noindent 
Fig. 15 -- Same as Figure 12 but for 4H47.

\newpage

\begin{table}[t]
\caption{Parameters for the progenitor models of SN 1993J}
\begin{center}
\begin{tabular}{cccccccc}
\tableline
Name & $M_{\rm env}/M_{\odot}$ & $M_{\rm ej}/M_{\odot}$ 
& $M_{\rm Ni}/M_{\odot}$ & $R_{i}/R_{\odot}$ & 
$L/L_{\odot}$ & $Y$ &  $E_{\rm exp}$( $10^{51}$ erg) \\
\tableline
3H11 & 0.11 & 2.06 & 0.1 & 450 & $4.1\times 10^{4}$ & 0.54 & 1.0 \\
3H41 & 0.41 & 2.36 & 0.1 & 450 & $4.1\times 10^{4}$ & 0.54 & 1.0 \\
4H13 & 0.13 & 2.78 & 0.1 & 580 & $6.8\times 10^{4}$ & 0.56 & 1.0 \\
4H47 & 0.47 & 3.12 & 0.1 & 350 & $6.8\times 10^{4}$ & 0.79 & 1.0 \\
\tableline
\end{tabular}
\end{center}
\end{table}

\begin{center}
$M_{\rm env}$: envelope mass, $M_{\rm ej}$: total
ejecta mass, $M_{\rm Ni}$: \ni~ mass. $R_{i}$: initial radii, 

$L$: presupernova luminosity, $Y$: helium mass fraction in
the envelope.
\end{center}

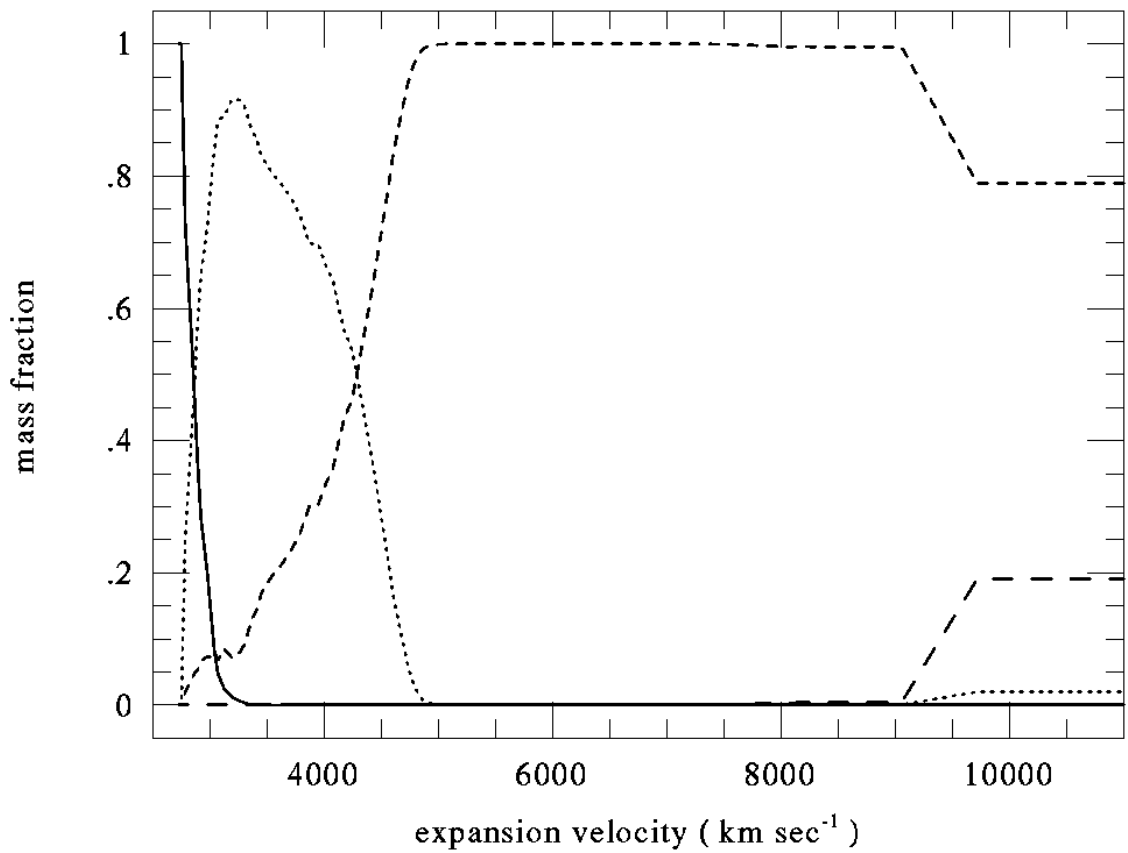
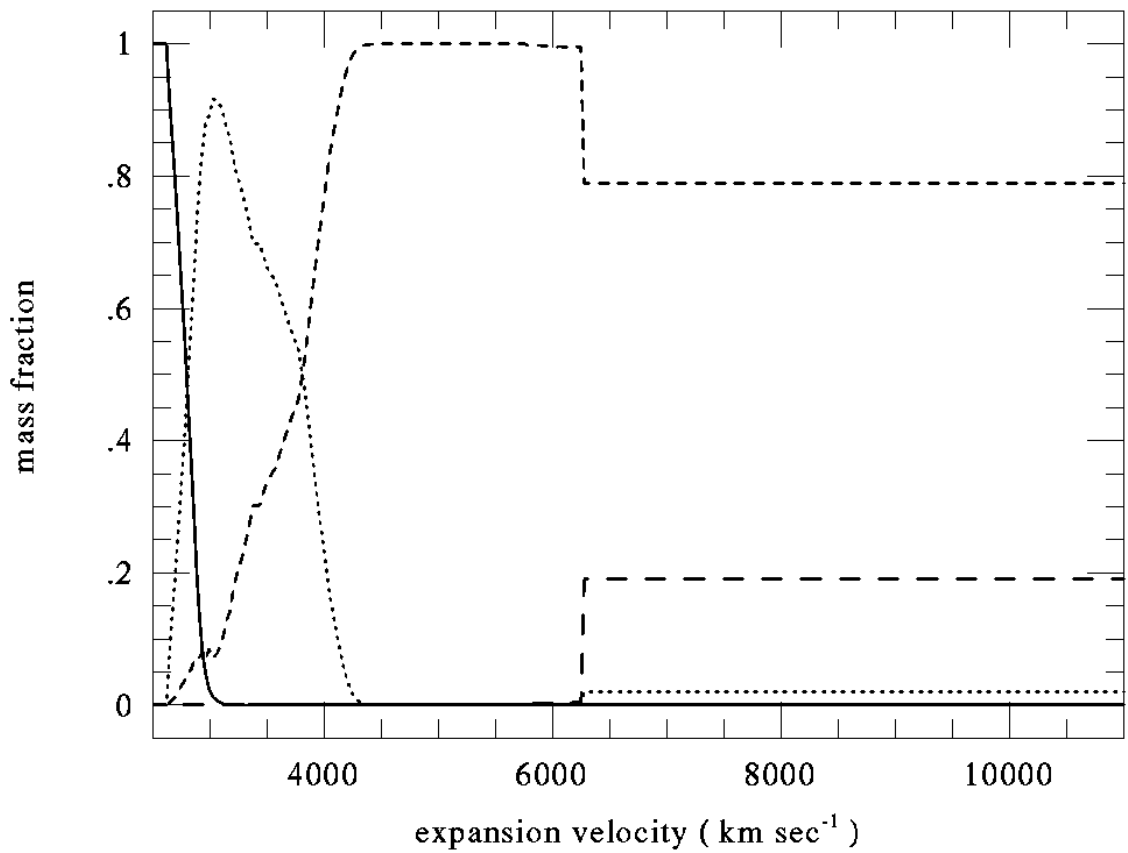
\end{document}